\documentclass{ws-procs10x7}

\begin{document}

\title{Hard diffraction --- 20 years later}

\author{Gunnar Ingelman}

\address{High Energy Physics, Uppsala University, Box 535, SE-75121 Uppsala, Sweden\\E-mail: gunnar.ingelman@tsl.uu.se}

\twocolumn[\maketitle\abstract{The idea of diffractive processes with a hard scale involved, to resolve the underlying parton dynamics, was published 1985 and experimentally verified 1988. Today hard diffraction is an active research field with high-quality data and new theoretical models. The trend from Regge-based pomeron models to QCD-based parton level models has given insights on QCD dynamics involving perturbative gluon exchange mechanisms, including the predicted BFKL-dynamics, as well as novel ideas on non-perturbative colour fields and their interactions. Extrapolations to the LHC include the interesting possibility of diffractive Higgs production. 
}]

\section{Introduction}\label{sec:introduction}
`Gaps -- in my understanding after 20 years' could have been an appropriate title on this talk, which will focus on today's understanding of hard diffraction based on models that include both hard perturbative and soft non-perturbative QCD. I will start, however, with some of the `historical' milestones that have established this new research field. Recent data have revealed problems with models based on the pomeron in `good old' Regge phenomenology. The key issue is whether the pomeron is a part of the proton wave function or diffraction is an effect of the scattering process. The latter seems more appropriate in today's QCD-based models. 
  
Recently, a new kind of `hard gap' events, having a large momentum transfer across the gap, have been observed in terms of a rapidity gap between two high-p$_\perp$ jets in $p\bar{p}$ and diffractive vector meson production at large momentum transfer in $\gamma p$. This has given the first real evidence for the BFKL-dynamics predicted since long by QCD. 

Thus, hard diffraction provides a `QCD laboratory' where several aspects of QCD dynamics can be investigated. Remember that QCD, in particular in its non-perturbative domain, still has major unsolved problems. 

Based on our current theoretical understanding, interesting predictions are made for hard diffraction in future experiments, {\it e.g.}\ at the LHC. Here, much interest is presently focused on the possibility to produce the Higgs boson in diffractive events, which may even provide a potential for Higgs discovery! 

\section{`Historical' milestones}\label{sec:historical}
\begin{figure}
\begin{center}
\epsfig{file=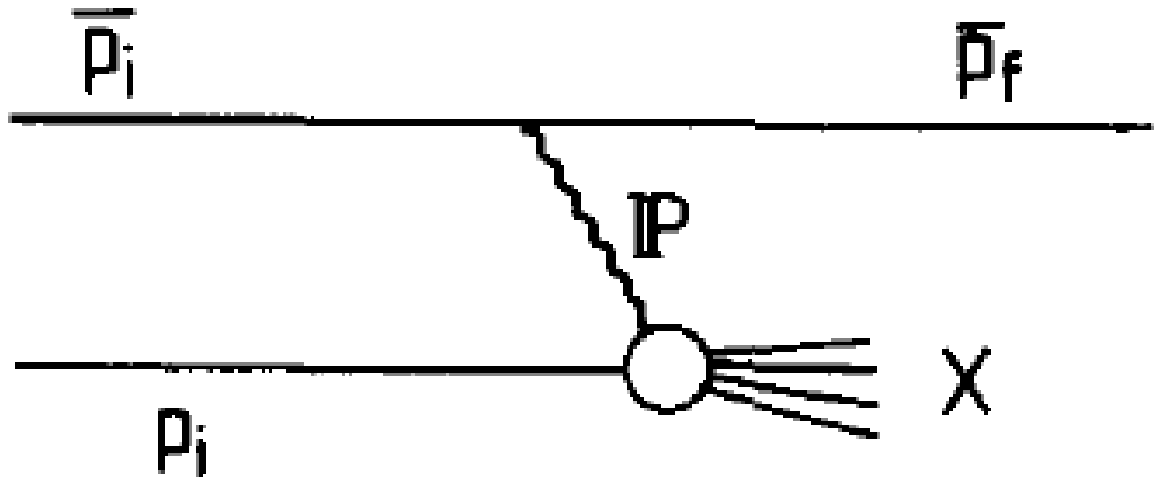,width=30mm}
\hspace*{1mm}
\epsfig{file=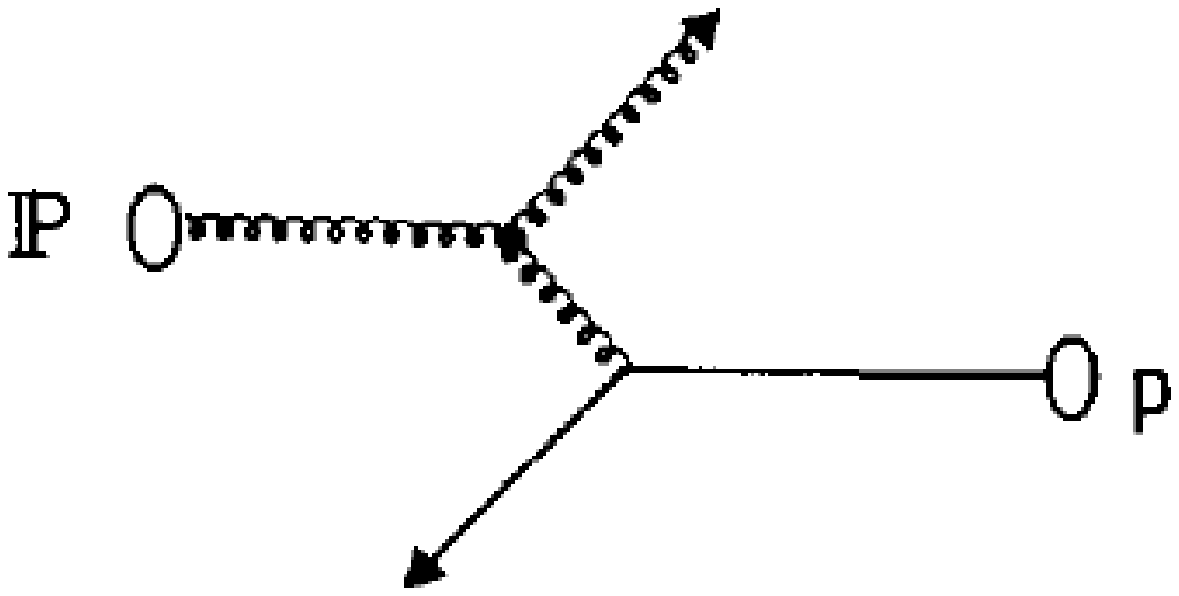,width=30mm}\\
\epsfig{file=UA8-lego.ps,bbllx=25pt,bblly=450pt,bburx=535,bbury=700,clip=, width=50mm}
\end{center}
\caption{(a) Single diffractive $p\bar{p}$ scattering via pomeron exchange giving a leading beam particle separated by a rapidity gap to the $X$-system. (b) Hard parton level scattering in the $X$-system producing high-$p_\perp$ jets. (c) Jets in the central calorimeter of an event triggered by a leading proton in the UA8 Roman pot detectors giving the discovery of hard diffraction.}
\label{fig:is-ua8}
\end{figure}

Going back to `ancient' pre-QCD history, Regge theory provided a phenomenology for total and diffractive cross-sections with a dominant contribution from the exchange of a pomeron with vacuum quantum numbers. Hadronic scattering events were then classified as elastic, single and double diffraction, double pomeron exchange and totally inelastic depending on the observable distribution of final state particles in rapidity.\footnote{Rapidity $y=\frac{1}{2}\ln\frac{E+p_z}{E-p_z} \approx -\ln \tan\frac{\theta}{2}=\eta$ pseudo-rapidity for a particle with $(E,\vec{p}_\perp,p_z)$ and polar angle $\theta$ to beam axis $z$.} For example, single diffraction is then characterised by a leading proton (or other beam hadron) separated by a large rapidity gap ({\it i.e.}\ without particles) to the $X$-system of final state particles (Fig.~\ref{fig:is-ua8}a). One should note that in this Regge approach, there is no hard scale involved. In particular, there is no large momentum transfer across the gap which may therefore be called a `soft gap'.

The basic new idea introduced 20 years ago by myself and Peter Schlein\cite{Ingelman:1984ns}, was to consider a hard scale in the $X$-system in order to resolve an underlying parton level interaction and thereby be able to investigate the process in a modern QCD-based framework. We formulated this in a model with an effective pomeron flux in the proton, $f_{I\!\!P /p}(x_{I\!\!P},t)$, and parton distributions in the pomeron $f_{q,g/I\!\!P}(z,Q^2)$, such that cross-sections for hard diffractive processes could be calculated from the convolution 
\begin{equation}
d\sigma \sim f_{I\!\!P /p} \; f_{q,g/I\!\!P} \; f_{q,g/p} \; d\hat{\sigma}_{\rm{pert.\ QCD}}
\label{eq:is-convolution}
\end{equation}
of these functions with a perturbative QCD cross-section for a hard parton level process (Fig.~\ref{fig:is-ua8}ab). This enabled predictions of diffractive jet production at the CERN $p\bar{p}$ collider and also of diffractive deep inelastic scattering. Although this seems quite natural in today's QCD language, it was rather controversial at the time. 

It was therefore an important break-through when the UA8 experiment at the CERN $p\bar{p}$ collider actually discovered\cite{Bonino:1988ae} hard diffraction by triggering on a leading proton in their Roman pot detectors and finding jets in the UA2 central calorimeter (Fig.~\ref{fig:is-ua8}c) in basic agreement with our model implemented in a `Lund Monte Carlo' event generator. These observed jets had normal jet properties and by investigating their longitudinal momentum distribution one could infer that the partons in the pomeron have rather a hard distribution, $f_{q,g/I\!\!P}(z)\sim z(1-z)$ and also indications\cite{Brandt:1992zu} of a superhard component $\sim \delta(1-z)$.

\begin{figure}
\begin{center}
\epsfig{file=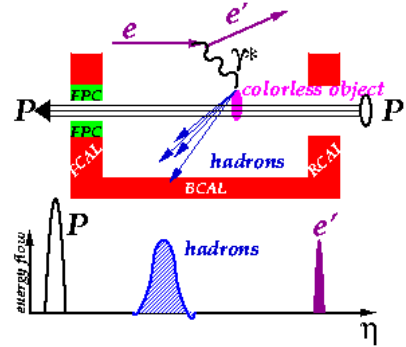,width=0.8\columnwidth}
\end{center}
\caption{Schematics of diffractive deep inelastic scattering in the ZEUS detector and the final state rapidity distribution.}
\vspace*{-5mm}
\label{fig:ddis-zeus}
\end{figure}

In spite of this discovery, hard diffraction was not fully recognised in the whole particle physics community. It was therefore a surprise to many when diffractive deep inelastic scattering (DDIS) was discovered by ZEUS\cite{ZEUS-DDIS} and H1\cite{H1-DDIS} at HERA in 1993. These events were quite spectacular with the whole forward detector empty (Fig.~\ref{fig:ddis-zeus}), {\it i.e.}\ a large rapidity gap as opposed to the abundant forward hadronic activity in normal DIS events. A surprisingly large fraction $\sim 10$\% 
of all DIS events were diffractive. Moreover, they showed the same $Q^2$ dependence and were not suppressed with increasing $Q^2$, demonstrating that DDIS is {\em not} a higher twist process but leading twist. 
 
\begin{figure}
\begin{center}
\epsfig{file=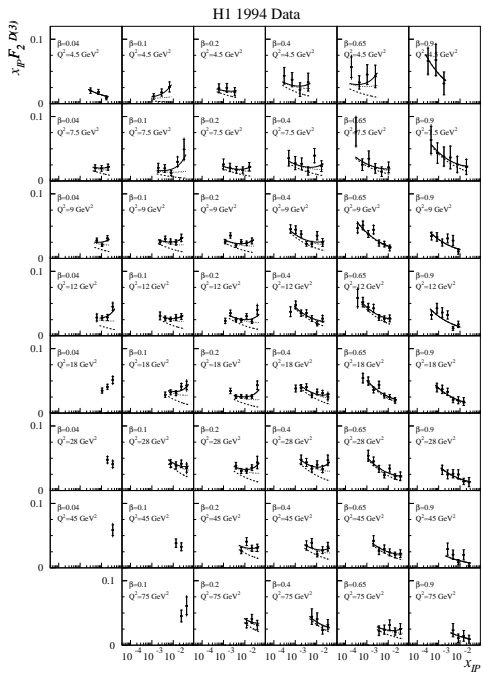,width=1.0\columnwidth}
\end{center}
\caption{H1 data\protect\cite{Adloff:1997sc} on the diffractive structure function $F_2^{D(3)}(x_{I\!\!P};\beta,Q^2)$ with fits based on the Regge models with pomeron and reggeon exhange.}
\label{fig:f2d3}
\end{figure}

The diffractive DIS cross-section can be written \cite{Ingelman:1992qf}
\begin{equation}
\frac{d\sigma}{dx\,dQ^2\,dx_{I\!\!P}\,dt}=\frac{2\pi\alpha^2}{xQ^4}
\left( 1+(1-y)^2\right) F_2^{D(4)} 
\label{eq:ddis-xsection}
\end{equation}
where fractional energy loss $x_{I\!\!P}$  and four-momentum transfer $t$  from the proton define the diffractive conditions. For most of the data, the leading proton is not observed and hence $t$ is effectively integrated out giving the structure function $F_2^{D(3)}(x_{I\!\!P},\beta,Q^2)$ which has then been obtained from rapidity gap events with high precision (Fig.~\ref{fig:f2d3}). The variables 
\begin{eqnarray}
\beta &=& \frac{-q^2}{2q\cdot (p_p-p_Y)} = \frac{Q^2}{Q^2+M_X^2-t} \\
x_{I\!\!P} &=& \frac{q\cdot (p_p-p_Y)}{q\cdot p_p} = \frac{Q^2+M_X^2-t }{Q^2+W^2-M_p^2} = \frac{x}{\beta} \nonumber
\label{eq:beta-x}
\end{eqnarray}
are model-independent invariants.
 
In $p\bar{p}$, UA8 has provided more information on diffractive jet production through analyses of such cross-sections\cite{ua8-analyses}. Several different diffractive hard scattering processes have been observed in $p\bar{p}$ at the Tevatron. Events with jets, $W$, $Z$, $b\bar{b}$ or $J/\psi$ have a large rapidity gap, and are thus diffractive, in about 1\% 
of the cases, as shown in Table~\ref{tab:diffr-tevatron}. This is an order of magnitude smaller relative rate of hard diffraction compared to the 10\%
in DIS at HERA. 

\begin{table}
\begin{center}
\begin{tabular}{llll}
\multicolumn{4}{c}{$R_{\mathrm{hard}} = \frac{1}{\sigma_{\mathrm{hard}}^{\mathrm{tot}}}
\int_{{x_F}_{\mathrm{min}}}^1 dx_F \, \frac{d\sigma_{\mathrm{hard}}}{dx_F}$}
\vspace*{2mm}\\
\hline
\hline
$R_{\mathrm{hard}} [\%
]$ & \multicolumn{2}{c}{Exp. observed} & SCI \\
\hline
dijets   & {\small CDF} & 0.75 $\pm$ 0.10           & 0.7  \\
W        & {\small CDF} & 1.15 $\pm$ 0.55           & 1.2  \\
W        & {\small D\O} & 1.08 $^{+0.21}_{-0.19}$   & 1.2  \\
$b\bar{b}$   & {\small CDF} & 0.62 $\pm$ 0.25           & 0.7  \\
Z        & {\small D\O} & 1.44 $^{+0.62}_{-0.54}$   & 1.0$^\star$ \\
$J/\psi$ & {\small CDF} & 1.45 $\pm$ 0.25           & 1.4$^\star$ \\
\hline
\hline
\end{tabular}
\caption{Tevatron data on the ratio in \% 
of diffractive hard processes to all such hard events, where diffraction is defined by a rapidity gap corresponding to a leading proton with large $x_F$. For comparison results of the SCI model discussed in section \ref{sec:qcd} ($^\star$ denote predictions in advance of data).}
\label{tab:diffr-tevatron}
\end{center}
\end{table}

\section{Pomeron approach}\label{sec:pomeron}
Using Regge factorisation, the diffractive DIS structure function can be written 
\begin{equation}
F_2^{D(4)}(x,Q^2,x_{I\!\!P},t)=f_{I\!\!P /p}(x_{I\!\!P},t) F_2^{I\!\!P}(\beta,Q^2)
\label{eq:f2d4}
\end{equation}
in terms of a pomeron flux and a pomeron structure function, where $x_{I\!\!P} \simeq p_{I\!\!P}/p_p$ is interpreted as the momentum fraction of the pomeron in the proton and $\beta \simeq p_{q,g}/p_{I\!\!P}$ is the momentum fraction of the parton in the pomeron. 

\begin{figure}[t]
\begin{center}
\epsfig{file=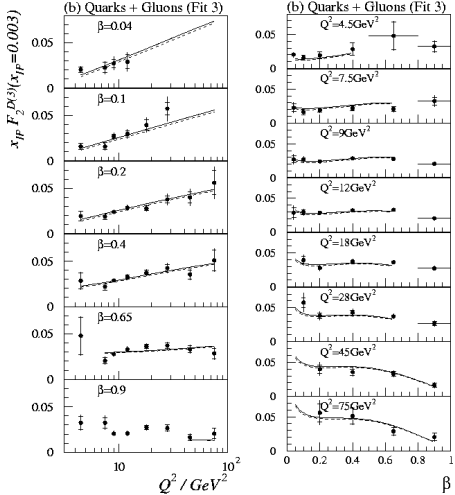,width=0.9\columnwidth}
\end{center}
\caption{H1 data\protect\cite{Adloff:1997sc} on the diffractive structure function $F_2^{D(2)}(\beta ,Q^2)$ with QCD fits.}
\label{fig:f2d2}
\end{figure}

Good fits with data can be obtained (Fig.~\ref{fig:f2d3}), provided that also a Reggeon exchange contribution is included. Factoring out the fitted $x_{I\!\!P}$ dependence, one obtains the diffractive structure function $F_2^{D(2)}(\beta,Q^2)$ (or $F_2^{I\!\!P}$) shown in Fig.~\ref{fig:f2d2}. The $Q^2$ dependence is rather weak and thus shows approximate scaling indicating scattering on point-like charges. There is, however, a weak $\log{Q^2}$ dependence which fits well with conventional perturbative QCD evolution. 

\begin{figure}
\begin{center}
\epsfig{file=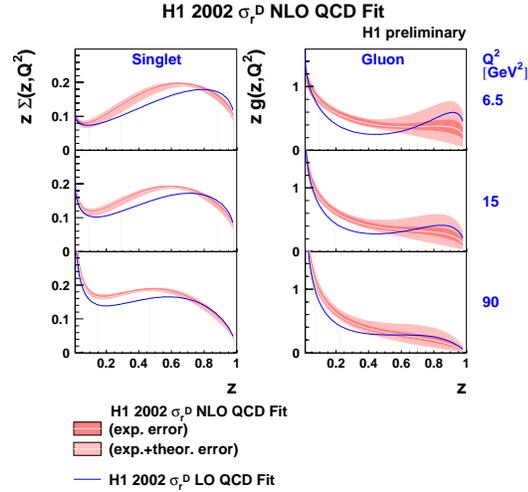,width=1.0\columnwidth}
\end{center}
\caption{Diffractive quark (singlet $z\Sigma (z,Q^2)$) and gluon ($zg(z,Q^2)$) momentum distributions (`in the pomeron') obtained\protect\cite{Adloff:1997sc} from a NLO DGLAP fit to H1 data on diffractive DIS.}
\label{fig:diffr-partons}
\end{figure}

The $\beta$ dependence is quite flat, which can be interpreted as hard parton distributions in the pomeron. This is borne out in a full next-to-leading order (NLO) QCD fit giving the parton distributions shown in Fig.~\ref{fig:diffr-partons}, which demonstrate the dominance of the gluon distribution. 

This framework can now be used to calculate various processes. In contributions\cite{HERA-diffr-D-dijets} to this conference, new HERA data on diffractive $D^\star$ and dijets are compared to calculations using these diffractive parton densities folded with the corresponding perturbative QCD matrix elements in NLO. For DIS the model calculation agree well with the data, both in absolute normalization and in the $Q^2$ and $p_\perp$ dependences. In photoproduction, the shapes of distributions agree with data, but the model normalization is too large by about a factor 2 (Table~\ref{tab:Dstar-dijets}). This fits with the theoretical knowledge that QCD factorization has been proven for diffraction in DIS\cite{Collins:1997sr}, but not in photoproduction ({\it i.e.}\ low $Q^2$) or hadronic interactions. Remember that the photon state $|\gamma\rangle = |\gamma\rangle _0 + |q\bar{q}(g)\rangle + |\rho\rangle \ldots$ has not only the direct component, but also hadronic components. 

\begin{figure}
\begin{center}
\includegraphics[bb=0 0 720 540,width=0.99\columnwidth]{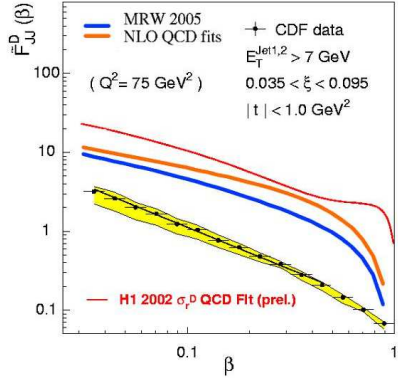}
\end{center}
\caption{D0 data on diffractive dijets at the Tevatron compared with model calculations based on diffractive parton densities from DIS at HERA.}
\label{fig:Fjj}
\end{figure}

Using this model with diffractive parton densities from HERA to calculate diffractive hard processes at the Tevatron, one obtains cross-sections which are an order of magnitude larger than observed, as shown in Fig.~\ref{fig:Fjj}. This problem can be cured by modifications of the model, in particular, by introducing some kind of damping\cite{pomeron-renormalisation} at high energies, such as a pomeron flux `renormalization'. It is, however, not clear whether this is the right way to get a proper understanding.

\begin{table}
\begin{center}
\begin{tabular}{|l|c|c|}
\hline 
$\frac{\sigma(\rm{data})}{\sigma(\rm{theory})}$ for &  H1   &   ZEUS  \\
\hline
$D^\star$ in diffr.\ DIS & $\sim 1$ & $\sim 1$ \\
dijets in diffr.\ DIS    & $\sim 1$ & $\sim 1$ \\
\hline 
$D^\star$ in diffr.\ $\gamma p$ & --- & $\sim 0.4$ \\
dijets in diffr.\ $\gamma p$    & $\sim 0.5$ & $\sim 0.5$ \\
\hline 
\end{tabular}
\caption{Ratio of HERA data on diffractive production of $D^\star$ and dijets to model calculation based on diffractive parton distributions and NLO perturbative QCD matrix elements.}
\label{tab:Dstar-dijets}
\end{center}
\end{table}

Thus, there are problems with the pomeron approach. The pomeron flux and the pomeron parton densities do not seem to be universal quantities. They cannot be separately well defined since only their product is experimentally measurable. Moreover, it may be improper to think of the pomeron as `emitted' from the proton, because the soft momentum transfer $t$ at the proton-pomeron vertex imply a long space-time scale such that they move together for an extended time which means that there should be some cross-talk between such strongly interacting objects. In order to investigate these problems, alternative approaches have been investigated where the pomeron is not in the initial state, {\it i.e.}\ not part of the proton wave function but an effect of the QCD dynamics of the scattering process. 

\section{QCD-based approaches}\label{sec:qcd}
A starting point can here be the standard hard perturbative interactions, since they should not be affected by the soft interactions. On the other hand we know that there should be plenty of soft interactions, below the cut-off $Q^2_0$ for perturbation theory, because $\alpha_s$ is then large giving a large interaction probability ({\it e.g.}\ unity for hadronisation). Soft colour exchange may then very well have a strong influence on the colour topology of the event and thereby on the final state via hadronisation. Ideally one would like to have a single model describing both diffractive gap events and non-diffractive events. 

\begin{figure}[t]
\begin{center}
\epsfig{file=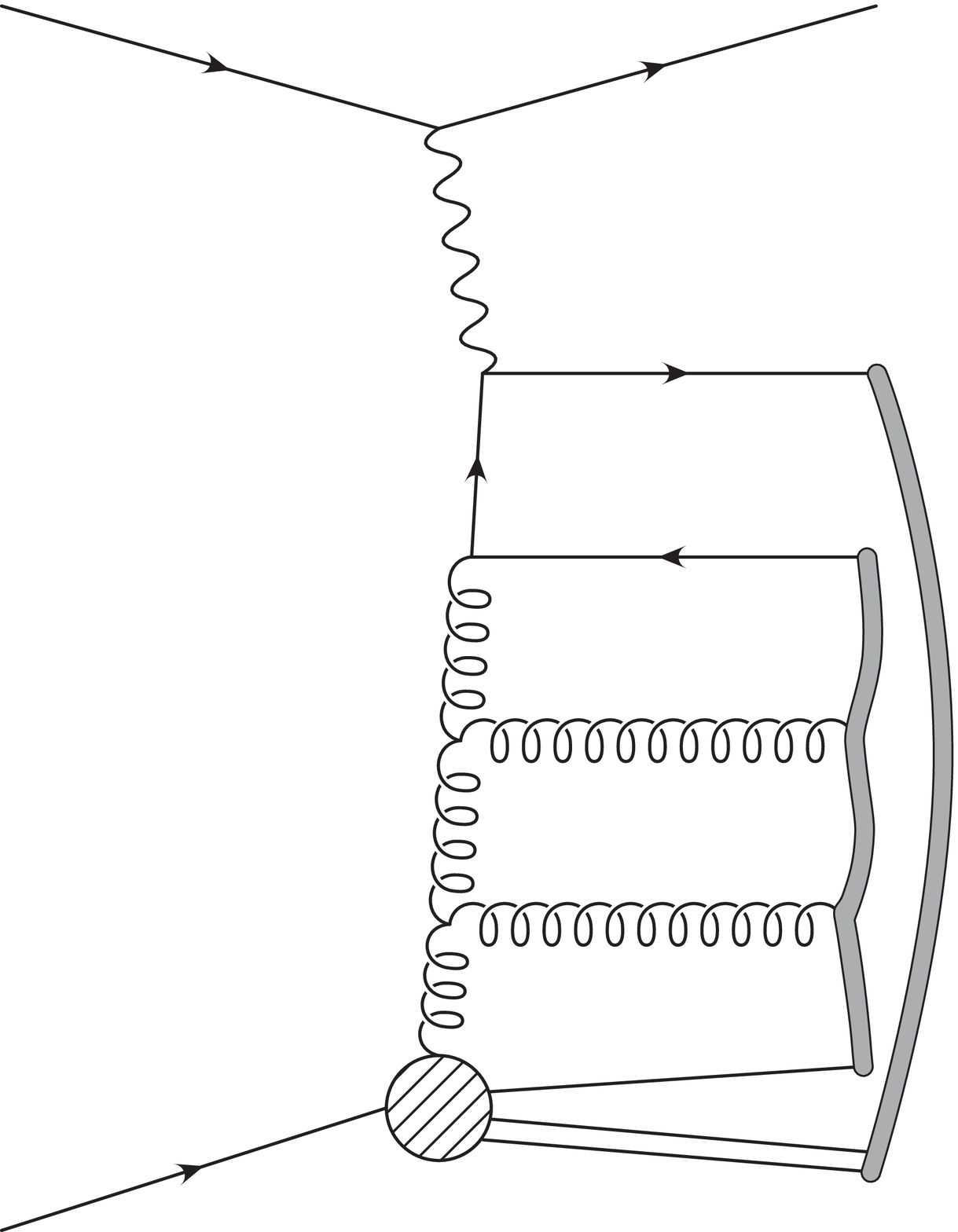,width=0.35\columnwidth}
\hspace*{5mm}
\epsfig{file=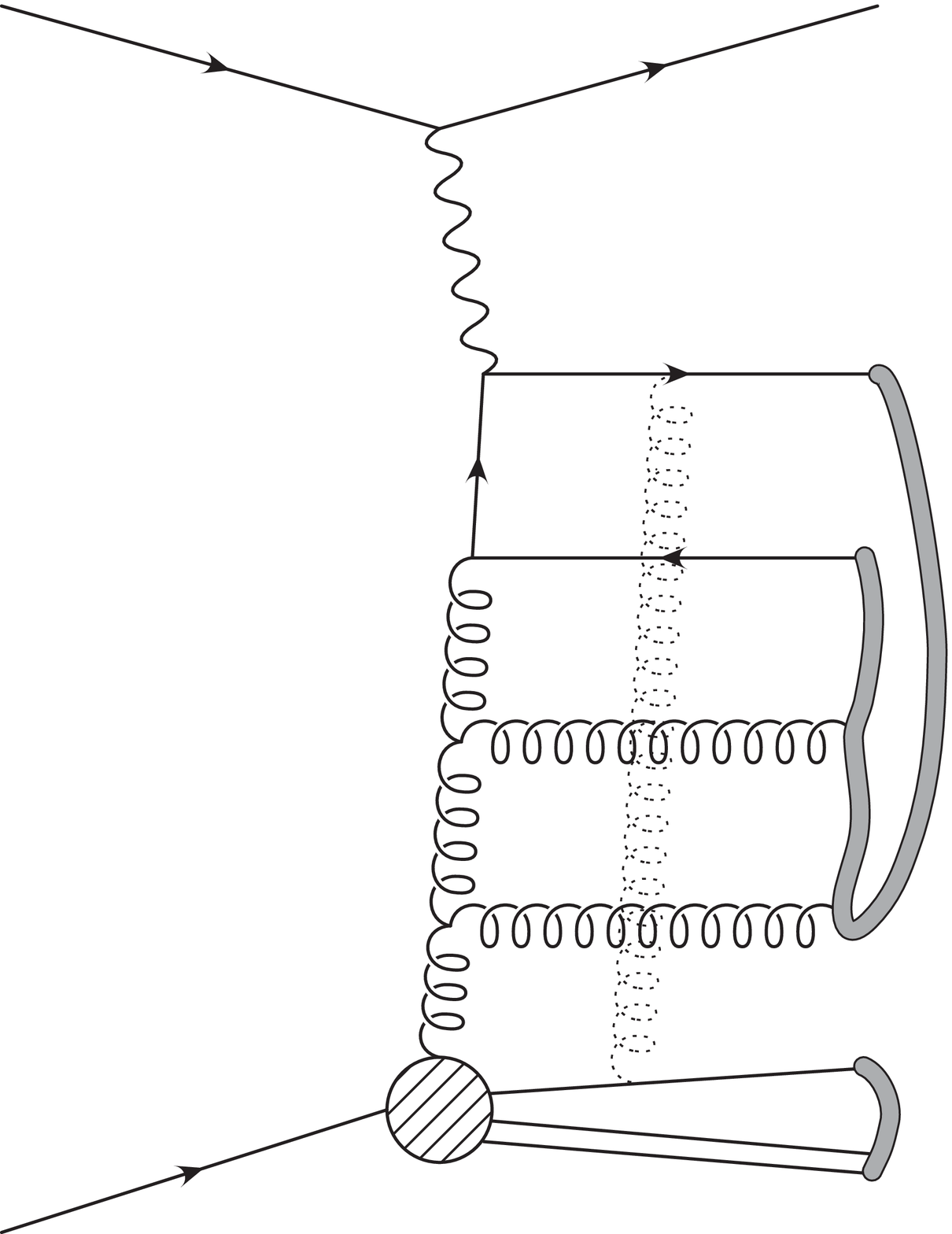,width=0.35\columnwidth}
\end{center}
\caption{Gluon-induced DIS at small $x$ with colour flux tube, or string, configuration in (a) the conventional Lund string model connection of partons and (b) after a soft colour-octet exchange (dashed gluon line) between the remnant and the hard scattering system resulting in a phase space region without a string leading to a rapidity gap after hadronisation.}
\label{fig:dis-strings}
\end{figure}

A simple, but phenomenologically successful attempt in this spirit is the soft colour interaction (SCI) model\cite{sci}. Consider DIS at small $x$, which is typically gluon-initiated leading to perturbative parton level processes as illustrated in Fig.~\ref{fig:dis-strings}. The colour order of the perturbative diagram has conventionally been used to define the topology of the resulting non-perturbative colour string fields between the proton remnant and the hard scattering system (Fig.~\ref{fig:dis-strings}a). 

Hadronisation, {\it e.g.}\ described by the Lund model\cite{Andersson:1983ia}, will then produce hadrons over the full rapidity region. One should remember, however, that we have proper theory to rely on only for the hard perturbative part of the event, which is separated from the soft dynamics in both the initial and the final parts by the QCD factorization theorem. This hard part is above a perturbative QCD cut-off $Q^2_0\sim 1 \mathrm{GeV}^2$ with an inverse giving a transverse size which is small compared to the proton diameter. Thus, the hard interactions can be viewed as being embedded in the colour field of the proton and hence one can consider interactions of the outgoing partons with this `background' field. Fig.~\ref{fig:dis-strings}b illustrates a soft gluon exchange that rearranges colour so that the hard scattering system becomes a colour singlet and the proton remnant another singlet. These systems hadronise independently of each other and are separated by a rapidity gap. The gap can be large because the primary gluon has a small momentum fraction $x_0$ given by the gluon density $g(x_0,Q^2_0)\sim x_0^{-\alpha}(1-x_0)^5$, leaving a large momentum fraction $(1-x_0)$ to the remnant which can form a leading proton.

The soft gluon exchange is non-perturbative and hence its probability cannot be calculated theoretically. The model, therefore, introduces a single parameter $P$ for the probability of exchanging such a soft gluon between any pair of partons, where one of them should be in the remnant representing the colour background field. Applying this on the partonic state, including remnants, in the Lund Monte Carlo generators {\sc Lepto} for $ep$ and {\sc Pythia} for hadronic interactions, {\it e.g.}\ $p\bar{p}$, leads to variations of the string topologies and thereby different final states after hadronisation. 

\begin{figure}[htb]
\begin{center}
\epsfig{file=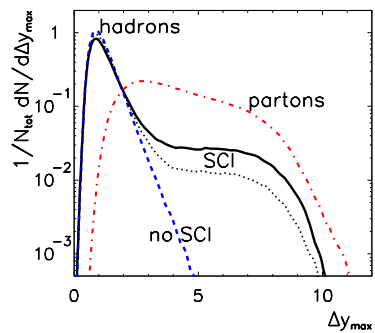,width=0.8\columnwidth}
\end{center}
\caption{Distribution of the size $\Delta y_{max}$ of the largest rapidity gap in DIS events at HERA simulated using {\sc Lepto} (standard small-$x$ dominated DIS event sample with $Q^2\ge 4$ GeV$^2$ and $x\ge 10^{-4}$).  The dashed-dotted curve represents the parton level obtained from hard, perturbative processes (matrix elements plus parton showers). The dashed curve is for the hadronic final state after standard Lund model hadronisation, whereas adding the Soft Color Interaction model results in the full curve. The dotted curve is when the SCI probability parameter $P$ has been lowered from its standard value 0.5 to 0.1.}
\label{fig:gap-size}
\end{figure}

The definition of diffraction through the {\em gap-size is a highly infrared sensitive observable}, as demonstrated in Fig.~\ref{fig:gap-size} for DIS at HERA. At the parton level, even after perturbative QCD parton showers, it is quite common to have large gaps. Hadronising the conventional string topology from the pQCD phase, leads to an exponential suppression with the gap-size, {\it i.e.}\ a huge non-perturbative hadronisation effect. Introducing the soft colour interactions causes a drastic effect on the hadron level result, with a gap-size distribution that is not exponentially suppressed but has the plateau characteristic for diffraction. The result of the SCI model is remarkably stable with respect to variation of the soft gluon exchange probability parameter, illustrating that the essential effect arises when allowing the possibility to rearrange the colour string topology. The gap events are in this approach nothing special, but a fluctuation in the colour topology of the event. 

Selecting the gap events in the Monte Carlo one can extract the diffractive structure function and the model (choosing $P\approx 0.5$) describes quite well\cite{Edin:1999jq} the main features of $F_2^{D(3)}(x_{I\!\!P},\beta,Q^2)$ observed at HERA (Fig.~\ref{fig:f2d3}). This is not bad for a one-parameter model! 

\begin{figure}[t]
\begin{center}
\epsfig{file=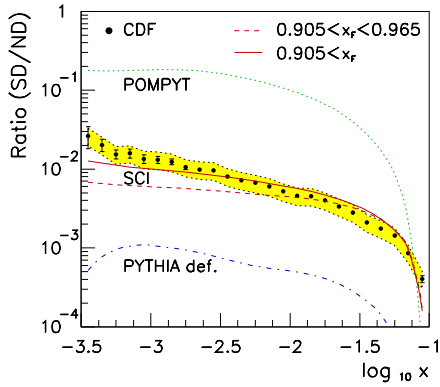,width=0.9\columnwidth}
\end{center}
\caption{Ratio of diffractive to non-diffractive dijet events versus momentum fraction $x$ of the interacting parton in $\bar{p}$. CDF data (with error band) compared to the {\sc Pompyt} pomeron model, default {\sc Pythia} and the SCI model.\protect\cite{Enberg:2001vq}}
\label{fig:dijets-x}
\end{figure}

By moving the SCI program code from {\sc Lepto} to {\sc Pythia}, exactly the same model can be applied to $p\bar{p}$ at the Tevatron. Using the value of the single parameter $P$ obtained at HERA, one obtains the correct overall rates of diffractive hard processes as observed at the Tevatron, see Table~\ref{tab:diffr-tevatron}. Differential distributions are also reproduced\cite{Enberg:2001vq} as exemplified in Fig.~\ref{fig:dijets-x}, which also demonstrates that the pomeron model is far above the data and {\sc Pythia} without the SCI mechanism is far below. The SCI model also reproduces the observed two-gap events (conventionally called double pomeron exchange) with a central hard scattering system\cite{Enberg:2001vq}.
 
\begin{figure}[t]
\begin{center}
\epsfig{file=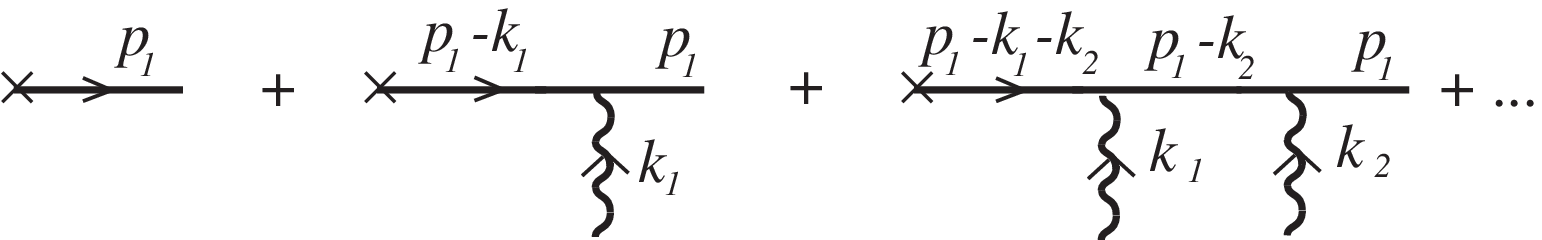,width=0.95\columnwidth}
\vspace*{3mm}\\
\epsfig{file=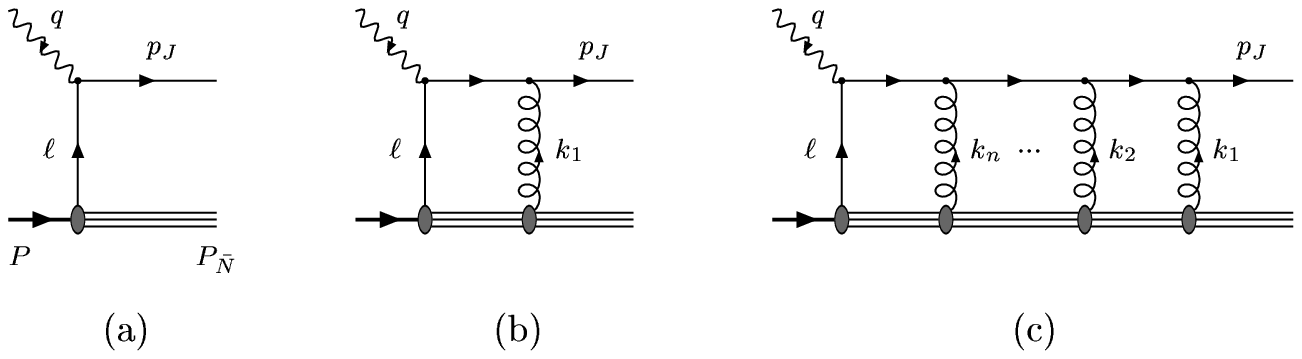,bbllx=76, bblly=101, bburx=447, bbury=168,clip=,width=0.95\columnwidth}
\end{center}
\caption{Diagram representation of Eq.~\ref{eq:wilson-expansion} with the scattered quark from a hard vertex (marked by the cross) and having 0,1,2\ldots rescatterings on the gauge field of the target (upper part) and its application in DIS (lower part).}
\label{fig:rescat-wilson}
\end{figure}

The phenomenological success of the SCI model indicates that it captures the most essential QCD dynamics responsible for gap formation. It is therefore interesting that recent theoretical developments provide a basis for this model. As is well known, the QCD factorization theorem separates the hard and soft dynamics and is the basis for the definition of the parton density functions (pdf) 
\begin{eqnarray}
f_{q/N}  & \sim & \int dx^- e^{-i x_B p^+ x^-/2} \langle \,N(p)\,|\, \bar\psi(x^-) 
\nonumber \\
 & & \gamma^+\, W[x^-;0] \, \psi(0)\,|\,N(p)\,\rangle_{x^+=0} \nonumber 
\label{eq:factorisation}
\end{eqnarray}
where the nucleon state sandwiches an operator including the Wilson line
$W[x^-;0] = {\rm P}\exp\left[ig\int_0^{x^-} dw^- A_a^+(0,w^-,0_\perp) t_a \right]$ which is a path-ordered exponential of gluon fields. The physical interpretation becomes transparent if one expands the exponential giving\cite{Brodsky:2002ue}
\begin{eqnarray}
 & & W[x^-;0] \sim 1 + g\int \frac{dk_1^+}{2\pi} \frac{\tilde A^+(k_1^+)}{k_1^+-i\varepsilon} + \\
 & & g^2 \int \frac{dk_1^+ dk_2^+}{(2\pi)^2}
\frac{\tilde A^+(k_1^+)\tilde A^+(k_2^+)}{(k_1^+ +k_2^+-i\varepsilon)
(k_2^+-i\varepsilon)} + \ldots \nonumber 
\label{eq:wilson-expansion}
\end{eqnarray}
with terms of different orders in the strong coupling $g$. As illustrated in Fig.~\ref{fig:rescat-wilson}, the first term is the scattered `bare' quark and the following terms corresponds to rescattering on the target colour field via 1,2\ldots gluons. This rescattering\cite{Brodsky:2002ue} has leading twist contributions for longitudinally polarised gluons, which are instantaneous in light-front time $x^+=t+z$ and occurs within Ioffe coherence length $\sim 1/m_px_{Bj}$ of the hard DIS interaction. 

\begin{figure}[h]
\begin{minipage}{\linewidth}
\centering\epsfig{file=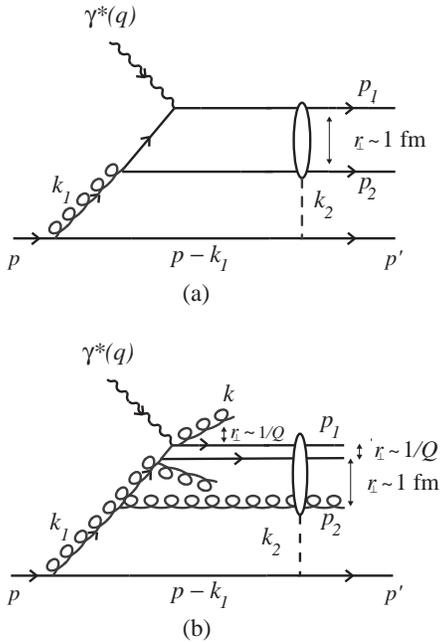,bbllx=0, bblly=0, bburx=186, bbury=274,clip=,width=0.85\columnwidth}
\caption{Low-order rescattering correction to DIS in 
the parton model frame where the virtual photon
momentum is along the negative $z$-axis with $q = (q^+,q^-,\vec
q_\perp) \simeq (-m_p x_B,2\nu,\vec 0)$ and the
target is at rest. The struck parton absorbs
nearly all the photon momentum giving $p_1 \simeq (0,2\nu,\vec
p_{1\perp})$ (aligned jet configuration). In (a) the virtual photon strikes a quark and the diffractive system is formed by the $q\bar{q}$ pair ($p_1,p_2$) which rescatters coherently from the target via `instantaneous' longitudinal ($A^+$) gluon exchange with momentum $k_2$. In (b) the $Q\bar{Q}$ quark pair which is produced in the $\gamma^* g \to Q\bar{Q}$ subprocess has a small transverse size $r_\perp \sim 1/Q$ and rescatters like a gluon. The diffractive system is then formed by the $(Q\bar{Q})\, g$ system. The possibility of hard gluon emission close to the photon vertex is indicated. Such radiation (labeled $k$) emerges at a short transverse distance from the struck parton and is not resolved in the rescattering.}
\label{fig:rescat-sci}
\end{minipage}
\end{figure}

This implies a rescattering of the scattered quark with the spectator system in DIS (Fig.~\ref{fig:rescat-wilson}). Although one can choose a gauge such that the scattered quark has no rescatterings, one can not `gauge away' all rescatterings with the spectator system\cite{Brodsky:2002ue}. The sum of the couplings to the $q\bar{q}$-system in Fig.~\ref{fig:rescat-sci}a gives the same result in any gauge and is equivalent to the colour dipole model in the target rest system (discussed below). Thus, there will always be such rescatterings and their effects are absorbed in the parton density functions obtained by fitting inclusive DIS data.

This has recently been used\cite{Brodsky:2004hi} as a basis for the SCI model. As illustrated in Fig.~\ref{fig:rescat-sci}a, a gluon from the proton splits into a $q\bar{q}$ pair that the photon couples to. Both the gluon and its splitting are mostly soft since this has higher probability ($g(x)$ and $\alpha_s$). The produced $q\bar{q}$ pair is therefore typically a large colour dipole that even a soft rescattering gluon can resolve and therefore interact with. The discussed instantaneous gluon exchange can then modify the colour topology before the string-fields are formed such that colour singlet systems separated in rapidity arise producing a gap in the final state after hadronisation as described by the SCI model. 

Similarly, the initial gluon may also split softly into a gluon pair followed by perturbative $g\to q\bar{q}$ giving a small $q\bar{q}$ pair (Fig.~\ref{fig:rescat-sci}b). Soft rescattering gluons can then not resolve the $q\bar{q}$, but can interact with the large-size $q\bar{q}$--$g$ colour octet dipole and turn that into a colour singlet system separated from the target remnant system that is also in a colour singlet state. Higher order perturbative emissions do not destroy the gap, since it occurs in the rapidity region of the hard system and not in the gap region. 

Moreover, the rescattering produces on-shell intermediate states having imaginary amplitudes\cite{Brodsky:2002ue}, which is a characteristic feature of diffraction. This theoretical framework implies the same $Q^2,x,W$ dependencies in both diffractive and non-diffractive DIS, in accordance with the observation at HERA. 

\begin{figure}
\begin{center}
\epsfig{file=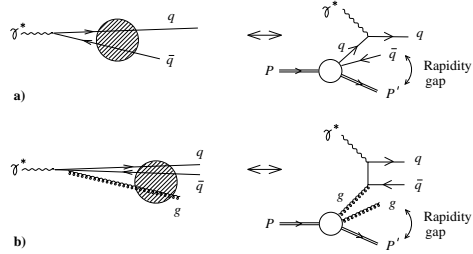,width=0.9\columnwidth}
\end{center}
\caption{Diffractive DIS in the semi-classical approach\protect\cite{buchmueller-hebecker} where the photon fluctuates into a $q\bar{q}$ or $q\bar{q}g$ system that interacts non-perturbatively with the proton colour field in the proton rest frame (left) and the corresponding Breit frame interpretations (right).}
\label{fig:buch-heb-graphs}
\end{figure}

In this approach the diffractive structure function is a convolution 
$F_2^D \sim g_p(x_{I\!\!P}) f(\beta )$ between the gluon density and the gluon splitting functions $f(\beta)\sim \beta^2+(1-\beta)^2$ for $g\to q\bar{q}$ and 
$f(\beta ) \sim (1-\beta(1-\beta))/(\beta(1-\beta)$ for $g\to gg$ (assuming that these perturbative expressions provide reasonable approximations also for soft splittings). Connecting to the pomeron language, this means that the gluon density replaces the pomeron flux and the gluon splitting functions the pomeron parton densities. One can note that the weak $\beta$ dependence observed in the diffractive structure function (Fig.~\ref{fig:f2d2}), here gets a natural explanation since $F_2^{I\!\!P}(\beta) = F_2^{D(4)}/f_{I\!\!P /p} \sim \frac{x}{\beta} F_2^{D(4)}(x)$, {\it i.e.}\ the applied factor $x/\beta$ essentially cancels the increase of the proton structure function $F_2$ at small $x$.

\begin{figure}[t]
\begin{minipage}{\linewidth}
\centering\epsfig{file=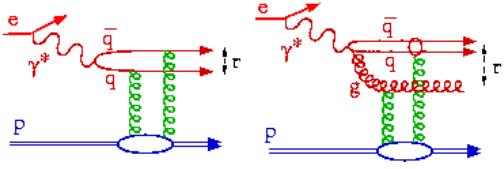,width=\linewidth}
\caption{Diffractive DIS in the model\protect\cite{Bartels:1998ea} based on two-gluon exchange between the non-perturbative proton and the perturbative fluctuation of the photon to $q\bar{q}$ and $q\bar{q}g$ colour dipoles.}
\label{fig:BEKW-graphs}
\end{minipage}
\end{figure}

Another approach, which has some similarities, is the so-called semi-classical approach\cite{buchmueller-hebecker}. The analysis is here made in the proton rest frame where the incoming photon fluctuates into a $q\bar{q}$ pair or a $q\bar{q}$--$g$ system that traverses the proton (Fig.~\ref{fig:buch-heb-graphs}). The soft interactions of these colour dipoles with the non-perturbative colour field of the proton is estimated using Wilson lines describing the interaction of the energetic partons with the soft colour field of the proton. The colour singlet exchange contribution to this process has been derived and shown to give leading twist diffraction when the dipole is large. This corresponds to a dipole having one soft parton (as in the aligned jet model), which is dominantly the gluon in Fig.~\ref{fig:buch-heb-graphs}b. One is thus testing the large distances in the proton colour field. This soft field cannot be calculated from first principles and therefore modelled involving parameters fitted to data. This theoretical approach is quite successful in describing the data (in Fig.~\ref{fig:f2d3}) on $F_2^{D(3)}(x_{I\!\!P};\beta,Q^2)$. 

Yet another approach starts from perturbative QCD and attempts to describe diffractive DIS as a two-gluon exchange\cite{Bartels:1998ea}. Again the photon fluctuates into a $q\bar{q}$ or a $q\bar{q}$--$g$ colour dipole. The upper part of the corresponding diagrams shown in Fig.~\ref{fig:BEKW-graphs} can be calculated in perturbative QCD giving essentially the $\beta$ and $Q^2$ dependencies of the diffractive structure function 
\begin{equation}
x_{I\!\!P}F_2^{D(3)}= F_{q\bar{q}}^T + F_{q\bar{q}g}^T + F_{q\bar{q}}^L
\end{equation}
with contributions of $q\bar{q}$ and $q\bar{q}g$ colour dipoles from  photons with transverse ($T$) and longitudinal ($L$) polarization given by
\begin{eqnarray}
F_{q\bar{q}}^T &=& A\left( \frac{x_0}{x_{I\!\!P}} \right)^{n_2} \beta (1-\beta)
\nonumber \\
F_{q\bar{q}g}^T &=& B\left( \frac{x_0}{x_{I\!\!P}} \right)^{n_2} \alpha_s \ln{(\frac{Q^2}{Q_0^2}+1)} (1-\beta)^{\gamma}
\nonumber \\
F_{q\bar{q}}^L &=& C\left( \frac{x_0}{x_{I\!\!P}} \right)^{n_4} \frac{Q_0^2}{Q^2} \left[ \ln{(\frac{Q^2}{4Q_0^2\beta}+1.75)}\right]^2  
\nonumber \\
  & &  \beta^3 (1-2\beta)^2
\nonumber 
\end{eqnarray}
where $n_{\tau} = n_{\tau 0}+n_{\tau 1}\ln{\left[ \ln{\frac{Q^2}{Q^2_0}}+1 \right] }$ for twist $\tau=2,4$.
\begin{figure}[t]
\begin{minipage}{\linewidth}
\centering\epsfig{file=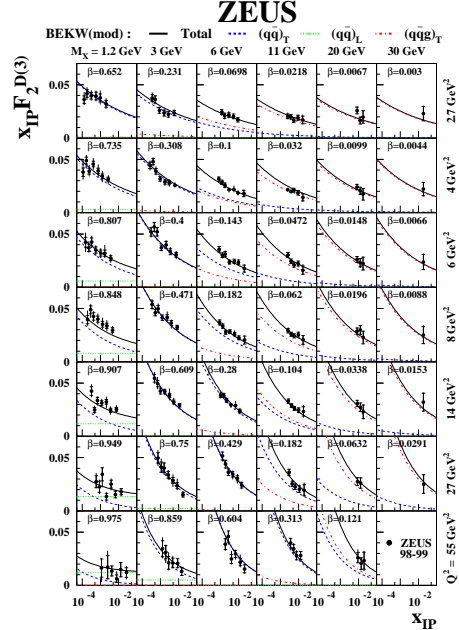,width=0.9\linewidth}
\caption{ZEUS data\protect\cite{Chekanov:2005vv} on the diffractive structure function $F_2^{D(3)}(x_{I\!\!P};\beta,Q^2)$ with fits of the model\protect\cite{Bartels:1998ea} for perturbative QCD two-gluon exchange.}
\label{fig:ZEUS-F2D-BEKW}
\end{minipage}
\end{figure}

The lower part of the diagram, with the connection to the proton, cannot be calculated perturbatively. This soft dynamics is introduced through a parameterisation where one fits the $x_{I\!\!P}$ dependence, which introduces parameters for the absolute normalization $A,B,C$ as well as $n_{\tau 0}, n_{\tau 1},\gamma$. The result\cite{Chekanov:2005vv} is a quite good fit to the data as shown in Fig.~\ref{fig:ZEUS-F2D-BEKW}. The different contributions from the two dipoles and photon polarizations are also shown, which provide interesting information on the QCD dynamics described by this approach. 

This perturbative two-gluon exchange mechanism is theoretically related to the processes in the following section. 

\section{Gaps between jets and BFKL}\label{sec:gaps-bfkl}
\begin{figure}
\centering\epsfig{file=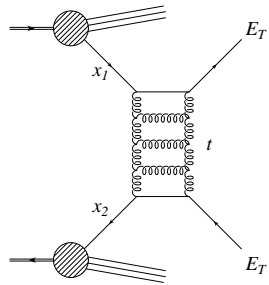,width=0.5\columnwidth}
\caption{Hard colour singlet exchange through a BFKL gluon ladder giving a rapidity gap between two high-$p_\perp$ jets.}
\label{fig:BFKL-ladder}
\end{figure}

In the processes discussed so far, the hard scale has not involved the gap itself since the leading proton has only been subject to a soft momentum transfer across the gap. A new milestone was therefore the observation\cite{D0-CDF-gap-jet-gap} at the Tevatron of events with a gap between two high-$p_\perp$ jets. This means that there is a large momentum transfer across the gap and perturbative QCD should therefore be applicable to understand the process. This is indeed possible by considering elastic parton-parton scattering via hard colour singlet exchange in terms of two gluons as illustrated in Fig.~\ref{fig:BFKL-ladder}. In the high energy limit $s/|t| \gg 1$, where the parton cms energy is much larger than the momentum transfer, the amplitude is dominated by terms $\sim [\alpha_s \, \ln (s/|t|)]^n$ where the smallness of $\alpha_s$ is compensated by the large logarithm. These terms must therefore be resummed leading to the famous BFKL equation describing the exchange of a whole gluon ladder (including virtual corrections and so-called reggeization of gluons). 

\begin{figure}[t]
\centering\epsfig{file=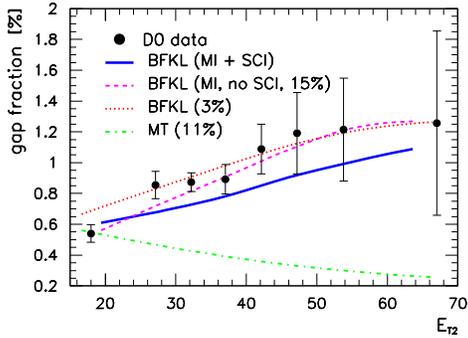,width=0.95\columnwidth}
\caption{Fraction of jet events having a rapidity gap in $|\eta|<1$ between the jets versus the second-highest jet-$E_T$. D0 data compared to the colour singlet exchange mechanism\protect\cite{Enberg:2001ev} based on the BFKL equation with non-leading corrections and with the underlying event treated in three ways: simple 3\% 
gap survival probability, {\sc Pythia}'s multiple interactions (MI) and hadronisation requiring a 15\% 
gap survival probability, MI plus soft colour interactions (SCI) and hadronisation with no need for an overall renormalisation factor. Also shown is the Mueller-Tang (MT) asymptotic result with a 11\%
gap survival probability.}
\label{fig:jet-gap-jet-ET}
\end{figure}

This somewhat complicated equation has been solved numerically\cite{Enberg:2001ev}, including also some non-leading corrections which turned out to be very important at the non-asymptotic energy of the Tevatron. This gave the matrix elements for an effective $2\to 2$ parton scattering process, which was implemented in the Lund Monte Carlo {\sc Pythia} such that parton showers and hadronisation could be added to generate complete events. This reproduces the data, both in shape and absolute normalization, which is not at all trivial, as demonstrated in Fig.~\ref{fig:jet-gap-jet-ET}. The non-leading corrections are needed since the asymptotic Mueller-Tang result has the wrong $E_T$ dependence. A free gap survival probability parameter, which in other models is introduced to get the correct overall normalization, is not needed in this approach. Amazingly, the correct gap rate results from the complete model including parton showers, parton multiple scattering and hadronisation through {\sc Pythia} together with the above discussed soft colour interaction model. The latter must be included, since the rescatterings are always present as explained above and without them an {\it ad hoc} 15\% 
gap survival probability factor would have to be introduced. 

This process of gaps between jets provides strong evidence for the BFKL dynamics as predicted since long by QCD, but which has so far been very hard to establish experimentally. 

\begin{figure}[t]
\begin{minipage}{0.95\linewidth}
\centering\epsfig{file=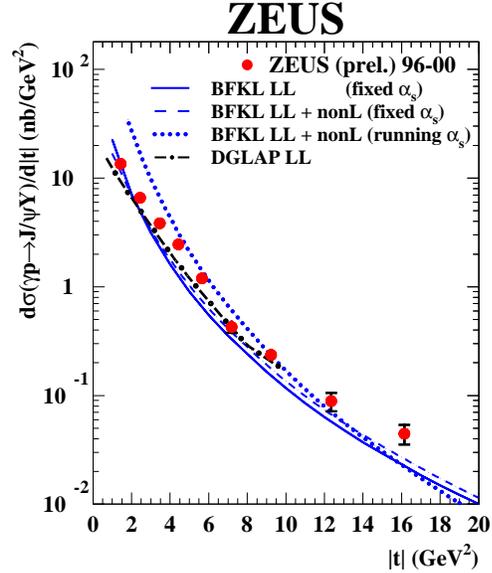,width=\columnwidth}\\
\caption{Differential cross-section $d\sigma/d|t|$ for the process $\gamma + p\to J/\psi + Y$. ZEUS data compared\protect\cite{ZEUS-diffr-psi-t} to BFKL model calculations using leading log (LL) with fixed $\alpha_s$, and including non-leading (non-L) corrections with fixed or running $\alpha_s$ as well as simple leading log DGLAP.}
\label{fig:ZEUS-diffr-psi-t}
\end{minipage}
\end{figure}

\begin{figure}[t]
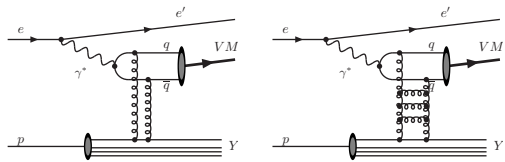

\begin{center}
\epsfig{file=ZEUS-diffr-psi-2gluon.epsi, bbllx=159pt,bblly=457pt,bburx=333pt,bbury=575pt,clip=,width=0.45\columnwidth}
\hspace*{2mm}
\epsfig{file=ZEUS-diffr-psi-BFKL.epsi, 
bbllx=159pt,bblly=457pt,bburx=333pt,bbury=575pt,clip=,width=0.45\columnwidth}\\
\end{center}
\caption{Diffractive vector meson production at large momentum transfer as described by perturbative QCD hard colour singlet exchange via two gluons and a gluon ladder in the BFKL framework\protect\cite{Enberg:2002zy}.}
\label{fig:ZEUS-diffr-psi-BFKL}
\end{figure}

Related to this is the new results from ZEUS\cite{ZEUS-diffr-psi-t} on the production of $J/\psi$ at large momentum transfer $t$ in photoproduction at HERA. The data, shown in Fig.~\ref{fig:ZEUS-diffr-psi-t}, agree well with perturbative QCD calculations\cite{Enberg:2002zy} (based on the hard scales $t$ and $m_{c\bar{c}}$) for two-gluon BFKL colour singlet exchange. As illustrated in Fig.~\ref{fig:ZEUS-diffr-psi-BFKL}, not only the simple two-gluon exchange is included, but also the full gluon ladder in either leading logarithm approximation or with non-leading corrections. Although the conventional DGLAP approximation can provide a good description in part of the $t$-region, in order to describe the full $t$-region and the energy dependence of this process one needs the BFKL formalism. Thus, this provides another evidence for BFKL dynamics.

\section{Future}\label{sec:future}
The discussions above illustrate that hard diffraction is a `laboratory' for QCD studies. This obviously includes small-$x$ parton dynamics and two- or multi-gluon exchange processes, such as gluon ladders and the BFKL equation. Moreover, high gluon densities lead to the concept of continuous colour fields and the interactions of a high energy parton traversing a colour field as described by the soft colour interaction model and the semiclassical approach discussed above. This has natural connections to the quark-gluon plasma and the understanding of the jet quenching phenomenon as an energy loss for a parton moving through the plasma colour field. 

Fundamental aspects of hadronisation also enters concerning gap formation and the production of leading particles from a small mass colour singlet beam remnant system. The latter is not only a problem in diffraction, but of more general interest, {\it e.g.}\ to understand how one should map a small-mass $c\bar{c}$ pair onto the discrete charmonium states\cite{Mariotto:2001sv}. 

We can still not exclude the possibility that the pomeron is some special kind of non-perturbative colour singlet gluonic system in the proton wave function. If so, this could be connected with the recently developed three-dimensional proton structure\cite{3D-proton-structure} $f_{q,g/p}(x,b,Q^2)$, where the quarks and gluons in the proton do not only have the normal $x,Q^2$ variables,  but also an impact parameter $b$ giving the transverse distance of the struck parton from the proton center as obtained by Fourier transformation of measured transverse momentum. Such analyses have led to speculations that the proton has a core of valence quarks surrounded by a cloud of sea partons. It is conceivable that such a cloud contains special gluon configurations that correspond to the pomeron. 

As long as QCD has important unsolved problems, diffraction will continue to be an interesting topic. This brings us to the LHC, where we will have all of the above diffractive processes but also some new ones. 

A topical and controversial issue currently is diffractive Higgs production at LHC. The first idea was here to exploit the fact that diffractive events are cleaner due to less hadronic activity and that it should therefore be easier to reconstruct a Higgs through its decay products in such an environment. This has even been considered as a Higgs discovery channel. At the Tevatron the cross-sections turns out to be very small due to the low energy available when the leading proton has only lost a small energy fraction\cite{diffr-H}. At the LHC, however, the cross-sections are quite reasonable, but Monte Carlo studies\cite{diffr-H} show that the events are not as clean as expected because the energy is so large that one can have a leading proton and a large gap (typically outside the detector) and still have plenty of energy to produce a lot of hadronic activity together with the Higgs.  

The exclusive process $pp\to p\, H\, p$ is, however, quite interesting. By measuring the protons one can here calculate the Higgs mass with missing mass techniques, without even looking at the central system. Therefore, plans are in progress for adding such leading proton spectrometers some hundred meters downstream in the LHC beam line ({\it e.g.}\ the `420 m' project). On the theoretical side there has been several model calculations for this exclusive process. The cross-section estimates vary by some orders of magnitude\cite{Khoze:2002py}, so some models must be substantially wrong.

\begin{figure}[t]
\begin{center}
\includegraphics[bb=0 0 720 540,width=0.9\columnwidth]{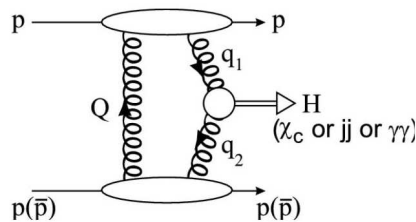}
\end{center}
\caption{Diagram for the exclusive process $pp\to p\, H\, p$, where the scales 
$M_H\gg Q \gg \Lambda_{QCD}$ motivate the use of perturbative QCD.}
\label{fig:exclusive-pHp}
\end{figure}

The most reliable state-of-the-art calculation is by the Durham group\cite{Martin:2005rz} based on the diagram in Fig.~\ref{fig:exclusive-pHp}. The basic process is calculated in perturbative QCD using Sudakov factors to include the requirement of no gluon radiation that would destroy the gap. There are, of course, also soft processes that might destroy the gap and these are taken into account by a non-trivial estimate of the gap survival probability. This gives a cross-section $\sigma \sim 3~fb$ for $M_H=120$ GeV at LHC giving $\sim 90$ events for $\int {\cal L} \sim 30~fb^{-1}$, which is certainly of experimental interest. At the Tevatron the cross-section is $\sigma \sim 0.2~fb$, which is too small to be of interest. Here, however, one can make important tests of this model calculation by instead of the Higgs consider similar exclusive production of smaller mass systems such as $\chi_c$, jet-jet and $\gamma\gamma$ which have larger cross-sections\cite{Martin:2005rz}.

\section{Conclusions}\label{sec:conclusions}
After 20 years of hard diffraction it is obvious that there has been great progress. Most importantly, this phenomenon has been discovered both in $p\bar{p}$ and $ep$ resulting in a lot of high-quality data, much more than could be presented here. The developments of theory and models have provided working phenomenological descriptions, but we do not have solid theory yet. Hard diffraction has become an important part of QCD research where, in particular, the interplay of hard and soft dynamics can be investigated.  

In the new QCD-based models, emphasized above, the pomeron is not a part of the proton wave function, but diffraction is an effect of the scattering process. Models based on interactions with a colour background field are here particularly intriguing, since they provide an interesting approach which avoids conceptual problems of pomeron-based models, such as the pomeron flux, but also provide a basis for constructing a common theoretical framework for all final states, diffractive gap events as well as non-diffractive events. 

But, there are still gaps in our understanding. This is not altogether bad, because it means that we have an interesting future first a few years at HERA and the Tevatron and then at the LHC. 



\clearpage
\section*{DISCUSSION}

\begin{description}
\item[Jon Butterworth] (UCL):\\
Is there a connection between the soft colour interactions which give rapidity gaps, and those which can form the $J/\psi$ in colour octet calculations?

\item[Gunnar Ingelman{\rm :}]
Yes, I think so. These are different theoretical techniques for describing the same, or similar, basic soft interactions. The colour octet model provides a nice theoretical framework with proper systematics, but one still need to fit uncalculable non-perturbative matrix elements to data. The SCI model started as a very simple phenomenological model and is now connected to QCD rescattering theory giving it more theoretical support. Also I think one should not see these different approaches, including the QCD-based models for diffraction that I have discussed, as excluding each other. It is not so that one is right and the others wrong. None is fully right, but they can be better or worse for describing and understanding different aspects of these poorly understood soft phenomena and all of them are likely to contribute to an improved understanding.

\item[Klaus M\"onig] (LAL/DESY):\\
If the Pomeron is in fact soft gluon exchange, shouldn't there be rapidity gaps also between quark and gluon jets at LEP?

\item[Gunnar Ingelman{\rm :}]
When using the SCI model for $e^+e^-$ we observed a very low rate of gaps, as is also the case in the data. This is due to the lack of an initial hadron giving a spectator system. The large gaps in $ep$ and $p\bar{p}$ depend on having a hadron remnant system taking a large fraction of the beam momentum when a soft gluon interacts. This remnant is then far away in rapidity from the rest and when it emerges as a colour singlet there will be large gap in rapidity after hadronisation. 

\item[Stephen L Olsen] (Hawaii):\\
At the B-factories we see an anomalously strong cross-section for exclusive 
$e^+e^- \to J/\psi \: \eta_c$ (at $\sqrt{s}=10.6$ GeV). This is as large a "gap" as possible at this energy. Can your methods be used to adress processes like
these?

\item[Gunnar Ingelman{\rm :}]
In principle yes, since the SCI model is implemented in the Lund Monte Carlos one could try it on anything that those Monte Carlos can simulate. However, there is a technical complication here when hadronising into an exclusive two-body final state, which the Lund hadronisation model is not constructed for and not suitable for. In spite of this, we have made recent progress in applying the model to exclusive $B$-meson decays, such as $B\to J/\psi K$, were the SCI mechanism increases the rate for such decay modes which are colour suppressed in the conventional theory. It may therefore be possible to apply it also for your process, but it cannot be obtained by just running the Monte Carlo straightforwardly---a dedicated study of the problem would be required.   

\end{description}

\end{document}